\documentclass[Physsubmission, Phys]{SciPost}

\hypersetup{
    colorlinks,
    linkcolor={red!50!black},
    citecolor={blue!50!black},
    urlcolor={blue!80!black}
}

\usepackage[bitstream-charter]{mathdesign}
\DeclareSymbolFont{usualmathcal}{OMS}{cmsy}{m}{n}
\DeclareSymbolFontAlphabet{\mathcal}{usualmathcal}

\begin{document}

\begin{center}{\Large \textbf{
Comparison of $pp$ and $p \bar{p}$ differential elastic cross sections 
and observation of the exchange of a Colorless $C$-odd gluonic compound\\
}}\end{center}

\begin{center}
Christophe Royon\textsuperscript{1}
\end{center}

\begin{center}
{\bf 1} The University of Kansas, Lawrence, USA
\\
* christophe.royon@ku.edu
\end{center}

\begin{center}
\today
\end{center}


\definecolor{palegray}{gray}{0.95}
\begin{center}
\colorbox{palegray}{
  \begin{tabular}{rr}
  \begin{minipage}{0.1\textwidth}
    \includegraphics[width=30mm]{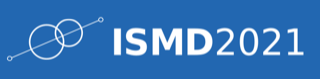}
  \end{minipage}
  &
  \begin{minipage}{0.75\textwidth}
    \begin{center}
    {\it 50th International Symposium on Multiparticle Dynamics}\\ {\it (ISMD2021)}\\
    {\it 12-16 July 2021} \\
    \doi{10.21468/SciPostPhysProc.?}\\
    \end{center}
  \end{minipage}
\end{tabular}
}
\end{center}

\section*{Abstract}
{\bf
We describe the discovery of the colorless $C$-odd gluonic compound, the odderon, by the D0 and TOTEM Collaborations by comparing elastic differential cross sections measured in $pp$ and $p \bar{p}$ interactions at high energies}

\vspace{10pt}
\noindent\rule{\textwidth}{1pt}
\tableofcontents\thispagestyle{fancy}
\noindent\rule{\textwidth}{1pt}
\vspace{10pt}

\section{Introduction: $pp$ and $p \bar{p}$ elastic data measuredly the D0 and TOTEM Collaborations}

\begin{figure}[h]
\centering
\includegraphics[width=0.7\textwidth]{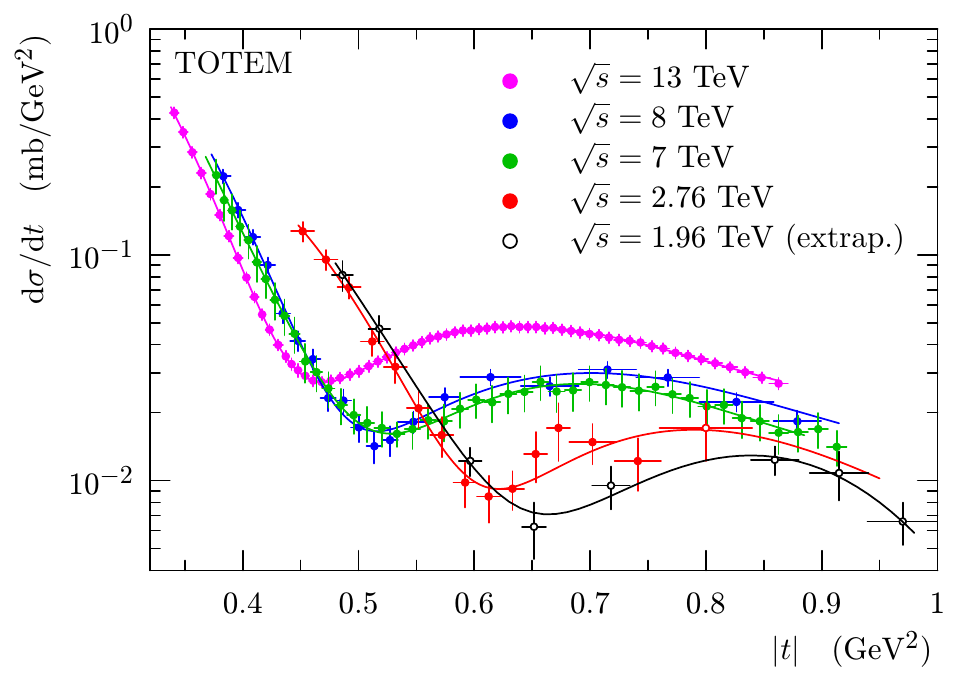}
\caption{TOTEM measured $pp$ elastic cross sections as a function of $|t|$ at 2.76, 7, 8, and 13 TeV (full circles), and the extrapolation (discussed in the text) to 1.96 TeV (empty circles).}
\label{fig1}
\end{figure}

Measurements of elastic scattering in $pp$ and $p \bar{p}$  interactions have been performed recently at the TeV scale at the Tevatron, Fermilab, USA and the LHC, CERN, Switzerland.  For this kind of intractions, the $p$ and $\bar{p}$ are intact and  scattered at small angles and no additional particle is produced. They can be detected and measured using dedicated detectors called roman pots, and vetoing on any particles that might be produced in addition to the intact $p$ or $\bar{p}$ in the central and forward detectors. Elastic events can be explained by the exchange of a colorless object, the pomeron or the odderon. 
Different experiments have been looking for the elusive odderon that was introduced about 50 years ago as a  singularity in the complex plane, located at $J = 1$ when $t = 0$ ($t$ is the transferred energy squared at the proton vertex) and which contributes to the odd crossing amplitude~\cite{nicolescu}. In terms of QCD, at least the odderon corresponds to an exchange of an odd number of gluons (dominated by 3 gluon exchanges) and the pomeron of an even number of gluons (dominated by 2 gluon exchanges). Observing differences between elastic $pp$ and $p \bar{p}$ interactions at high energies might lead to the discovery of the odderon~\cite{ourpaper}.

Differences were already observed between $pp$ and $p \bar{p}$ interactions at ISR energies~\cite{ISR} but this was not considered as evidence of the existence of the odderon. The difficulty is that, at low energies, elastic interactions can be due to exchange of pomerons and odderons, but also to reggeons, and mesons such as $\rho$, $\omega$ and $\phi$, and distinguishing between them becomes quickly model-dependent. The differences at ISR energies were interpreted as $\omega$ exchanges and not as a sign of the odderon. At high energies such as the Tevatron or the LHC, meson and reggeon exchanges become negligible (as can be seen from the smooth $t$-dependence of the cross section at 7, 8 and 13 TeV for instance) and this is why potential differences between $pp$ and $p \bar{p}$ interactions can be interpreted directly as evidence for the odderon.

The D0 collaboration measured elastic collisions in $p \bar{p}$ collisions at 1.96 TeV at the end of the Tevatron data taking using about 31 nb$^{-1}$ of data~\cite{d0cross} by measuring the $p$ and $\bar{p}$ in dedicated roman pots located at about 30 meters from the interaction point~\cite{FPD}.  Benefitting from the LHC running at different center-of-mass energies, the TOTEM collaboration also measured the elastic $pp$ differential cross section $d \sigma/dt$ at 2.76, 7, 8 and 13 TeV at the LHC~\cite{totemdata} using the same method as in D0, detecting the intact protons in roman pot detectors located at about 220 m from the interaction point~\cite{totem}. The measured elastic $pp$ differential cross sections by the TOTEM collaboration at different center-of-mass energies are shown in Fig.~\ref{fig1}~\cite{ourpaper}. Data always show the same features as a function of $|t|$, namely a decrease of the cross section that reaches a minimum, the dip, then an increase to reach a maximum, the bump, and again a decrease. $p \bar{p}$ elastic $d \sigma/dt$ as measured by the D0 Collaboration does not show the same features~\cite{d0cross}.  The idea is thus to analyze these differences in a quantitative way in order to find potential evidence for the odderon.

\begin{figure}[h]
\centering
\includegraphics[width=0.7\textwidth]{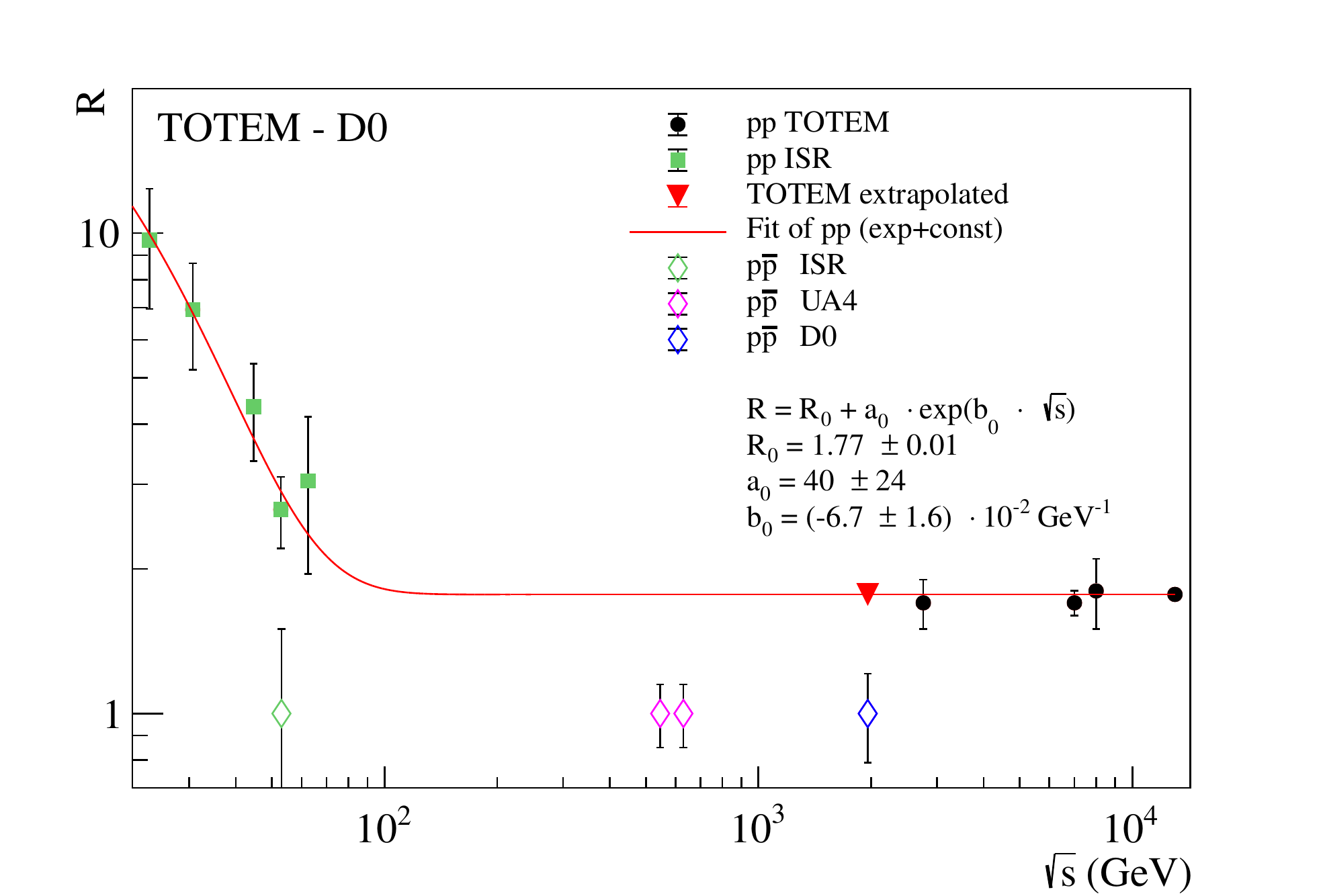}
\caption{$R$ Ratio of the elastic $d\sigma/dt$ cross sections at the bump and the dip as a function of $\sqrt{s}$ for $pp$ and $p \bar{p}$
interactions}
\label{fig2}
\end{figure}

The first obvious variable that we look at to distinguish between $pp$ and $p \bar{p}$ is the ratio $R$ of the elastic differential cross sections $d \sigma/dt$ at the bump and at the dip. Results are shown in  Fig.~\ref{fig2}. We notice the decrease of $R$ at ISR energies as a function of $\sqrt{s}$ for $pp$ interactions. At LHC energies, $R$ does not depend on $\sqrt{s}$ any longer and from this obervation, the extrapolated $R$ value from $pp$ TOTEM elastic data at 1.96 TeV is displayed as a red triangle. On the contrary, the $R$ values for $p \bar{p}$ from ISR, $Sp \bar{p}S$  and D0 measurements are  equal to 1.0, that leads to more than 3$\sigma$ difference between $pp$ and $p \bar{p}$ elastic data~\cite{ISR,SPS}.

\section{Prediction on $pp$ elastic $d\sigma/dt$ at 1.96 TeV from TOTEM data}

\begin{figure}[h]
\centering
\includegraphics[width=1.0\textwidth]{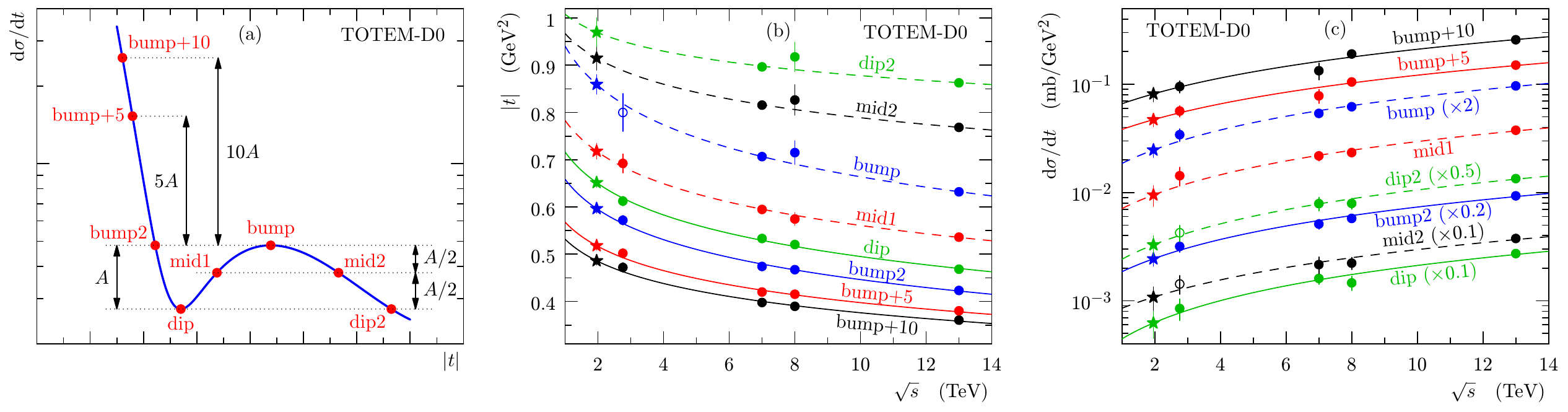}
\caption{$(a)$ Schematic definition of the characteristic points in the TOTEM differential cross section data.  $(b)$ and $(c)$ Characteristic points in $|t|$ and $d\sigma/dt$ from TOTEM measurements at 2.76, 7, 8, and 13 TeV (circles) as a function of $\sqrt{s}$ extrapolated to Tevatron center-of-mass energy (stars).  }
\label{fig3}
\end{figure}

The goal is first to characterize the behavior of the $pp$ elastic cross section at different center-of-mass energies as measured by TOTEM (namely 2.76, 7, 8 and 13 TeV) in order to compare with elastic $p \bar{p}$ interactions. The main point is that there is a bump and a dip for $pp$ interactions that are not observed in $p \bar{p}$ cross sections measurements. We thus define eight characteristic points around the dip and the bump that are characteristic of $pp$ elastic interactions. The definition of these eight characteristic points is shown in Fig.~\ref{fig3}, left. Using the TOTEM data, we measure the $t$ and $d \sigma/dt$ values of these characteristic points as a function of $\sqrt{s}$ as shown in Fig.~\ref{fig3}, middle and right. It is worth noting that  we choose the data points closest to the characteristic points in order to avoid model-dependent fits. A simple two-parameter fit in $t$ and $d \sigma/dt$ of these characteristic points as a function of $\sqrt{s}$ ($|t| = a \log (\sqrt{s}{\rm [TeV]}) + b$ and $(d\sigma/dt) = c \sqrt{s}~{\rm [TeV]} + d$) allows obtained the $t$ and $d \sigma/dt$ values of the characteristic points at the Tevatron energy $\sqrt{s}=1.96$ TeV.
Different kinds of parameterizations (with 2 or 3 parameters) were used and lead to similar results within 30\% of theu uncertainties. Let us notice that the 2.76 TeV data are crucial for this extrapolation procedure. Without these data, the range of extrapolation would be much larger (between 7 and 1.96 TeV). If it was possible to run the LHC at 1.96 TeV, we could get a direct comparison between elastic $pp$ and $p \bar{p}$ $d \sigma/dt$ but unfortunately, it is not easy to run the LHC at 2 TeV and there would be no acceptance in the dip and bump region, where we perform the comparison, with the present location of the roman pot detectors. A direct comparison would thus require deep modifications of the LHC machine which is obviously not possible at present with the goal of going towards high luminosity LHC.

In order to compare the extrapolated TOTEM elastic data with the D0 measurements, we need to compute the extrapolated elastic $pp$ $d \sigma/dt$ values in the same $|t|$ bins as for the D0 measurements. For this sake, we fit the reference points extrapolated to 1.96 TeV from TOTEM measurements  using a double exponential fit ($\chi^2=0.63$ per dof)
\begin{eqnarray}
h(t) = a_1 e^{-b_1 |t|^2 - c_1|t|}  
+ d_1 e^{-f_1 |t|^3 - g_1 |t|^2 -h_1 |t|}. \nonumber
\end{eqnarray}
This function is chosen for fitting purposes only and the two exponential terms cross around the dip, one rapidly falling and becoming 
negligible in the high $t$-range where the other term rises off the dip. Systematic uncertainties are evaluated from an ensemble of MC
 experiments in which the cross section values of the eight characteristic points are varied within their Gaussian uncertainties. The full covaraince matrix is used in order to take into account correlations between measured points. In addition, it is worth noting that
such formula leads also to a good description of TOTEM data in the dip/bump region at 2.76, 7, 8 and 13 TeV.

\begin{figure}[h]
\centering
\includegraphics[width=0.5\textwidth]{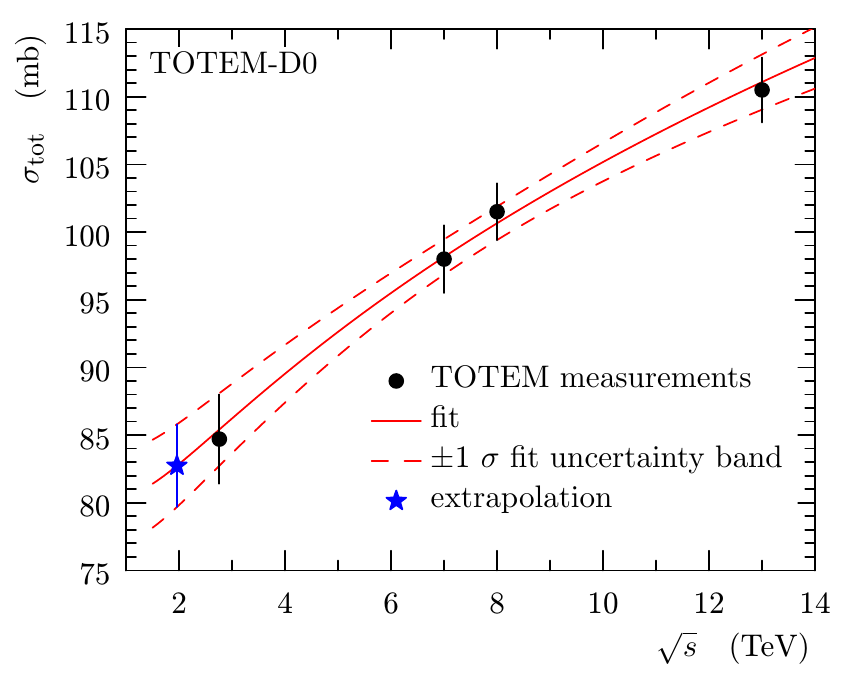}
\caption{Total cross section as a function of $\sqrt{s}$ measured by the TOTEM collaboration extrapolated (blue star) to the Tevatron center-of-mass energy.}
\label{fig4}
\end{figure}

The last step before comparing the TOTEM extrapolated data with the D0 $p \bar{p}$ measurements, it is needed to take into account the differences in normalization between both experiments. For instance, the D0 data show a fully correlated  absolute uncertainty of 14.5\% due to luminosity uncertainties. 
We thus adjust TOTEM and D0 data sets to have the same cross sections at the optical point (OP)
$d\sigma/dt(t = 0)$ (OP cross sections are expected
to be equal if there are only C-even exchanges and are different by at most 2\% in case of maximal odderon models which is taken as an additional systematic uncertainty).
The first step  is to predict the $pp$ total  cross section at 1.96 TeV from a fit to TOTEM data at higher center-o-mass energies as shown in Fig.~\ref{fig4},  $\sigma_{tot}=$82.7 $\pm$ 3.1 mb (with a $\chi^2=$0.27). We then obtain the value of $d\sigma/dt(t=0)$ at the OP using
\begin{eqnarray}
\sigma_{tot}^2 = \frac{16 \pi (\hbar c)^2}{1+\rho^2}  \left( \frac{d \sigma}{dt} \right)_{t=0} \nonumber
\end{eqnarray}
which leads to a TOTEM $d\sigma/dt(t=0)$ at the OP of 357.1 $ \pm$ 26.4 mb/GeV$^2$.
The D0 Collaboration measured the optical point of $d \sigma/dt$ at small $t$ to be 341$\pm$48 mb/GeV$^2$ and we thus rescale the TOTEM data by the ratio of these two numbers, namely 0.954 $\pm$ 0.071.
It is important to note that we do
not claim that we performed a measurement of $d\sigma/dt$ at
the OP at $t = 0$ (it would require additional measurements  closer to $t = 0$), but
we use the two extrapolations simply in order to obtain
a common and somewhat arbitrary normalization point.

\section{Comparison between the elastic $d\sigma/dt$ measurements from D0 and the extrapolated TOTEM data and the odderon discovery}

\begin{figure}[h]
\centering
\includegraphics[width=0.7\textwidth]{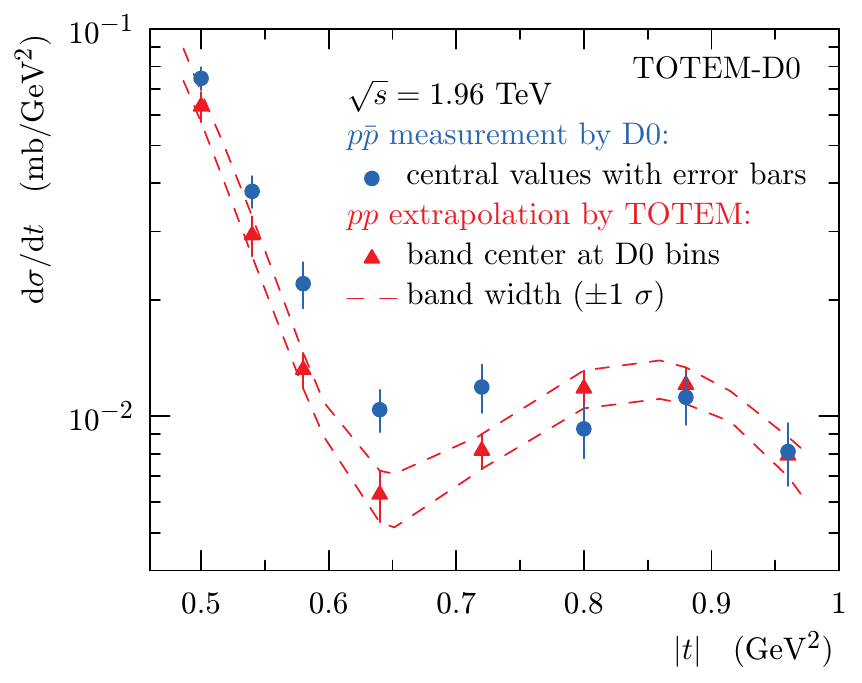}
\caption{Comparison between the D0 $p\bar{p}$ measurement
at 1.96 TeV and the extrapolated TOTEM $pp$ cross section, rescaled to match the OP of
the D0 measurement.  The dashed lines show the
1$\sigma$ uncertainty band on the extrapolated $pp$ cross section.}
\label{fig5}
\end{figure}

The comparison between the elastic $pp$ and extrapolated $p \bar{p}$ from the D0 and TOTEM collaborations is shown in Fig.~\ref{fig5}. The comparison is obviously done in the kinematical domain in $t$ where we have common data between D0 and TOTEM in order to be model-independent. Differences between both measurements are observed in the region 0.5 to 0,8 GeV$^2$ in $t$. In order to evaluate precisely the discrepancy, we perform a $\chi^2$ test
\begin{eqnarray}
\chi^2 = \Sigma_{i,j} [(T_i-D_i)C_{ij}^{-1} (T_j-D_j)] + \frac{(A-A_0)^2}{\sigma_A^2} + \frac{(B-B_0)^2}{\sigma_B^2}  \nonumber
\end{eqnarray}
where $T_j$ and $D_j$ are the $j^{th}$ $d\sigma/dt$ values for TOTEM and D0, $C_{ij}$ the covariance matrix, $A$ ($B$) the nuisance parameters  for scale (slope) with $A_0$ ($B_0$) their nominal values.
Slopes are constrained to their measured values ($pp$ to $p \bar{p}$ integrated elastic cross section ratio (dominated by the exponential part) becomes 1 in the limit $\sqrt{s} \rightarrow \infty$ which means similar slopes at small $|t|$ as observed in data).
The $\chi^2$ test uses the difference of the integrated cross section in the examined
$|t|$-range with its fully correlated uncertainty, and the experimental and extrapolated points with
their covariance matrices.
Given the constraints on the OP normalization and logarithmic slopes of the elastic cross sections, the $\chi^2$ test with six degrees of freedom yields the  $p$-value of 0.00061, corresponding to a significance of 3.4$\sigma$.

We can now combine this result with previous measurements performed by the TOTEM collaboration on the
total cross section and the $\rho$ parameter, the ratio of the real to imaginary part of the nuclear elastic amplitude at $t = 0$~\cite{rho}. This measurement was performed using data in the Coulomb-Nuclear interference region at very low $t$ (10$^{-3}$-10$^{-4}$ GeV$^2$) corresponding to data taking at $\beta^*=$2.5 km, and different detectors with respect to the data used in the D0/TOTEM comparison. Both results can thus be combined since they are independent.
Using low $|t|$ data in the Coulomb-nuclear interference region, $\rho$ was measured at 13 TeV to be $0.09 \pm 0.01$~\cite{rho}.
The combination of the measured $\rho$ and $\sigma_{tot}$ values are not compatible with any set of models without odderon exchange, leading to a 3.4 to 4.6$\sigma$ significance for the odderon.
When combined with the $\rho$ and total cross section result at 13 TeV, the total significance for the odderon is in the range 5.2 to 5.7$\sigma$ and thus constitutes the first experimental observation of the odderon.

\section{Conclusion}

We analyzed the differences between elastic $pp$ and $p \bar{p}$ interactions at 1.96 TeV by comparing the measurements of the D0 collaboration and the extrapolation of the TOTEM measurements at 2.76, 7, 8 and 13 TeV.
$pp$ and $p\bar{p}$ cross sections differ
with a significance of 3.4$\sigma$ in a model-independent way and thus provides evidence that the Colorless $C$-odd gluonic compound,  i.e. the odderon,
is needed to explain elastic scattering at high energies.
When combined with the $\rho$ and total cross section result at 13 TeV from the TOTEM Collaboration, the significance is in the range 5.2 to 5.7$\sigma$ and thus constitutes the first experimental observation of the odderon, which represents a  major discovery at CERN and Tevatron~\cite{ourpaper,press}.



\nolinenumbers

\end{document}